\documentclass[journal]{IEEEtran}
\usepackage{amsmath,amsfonts}
\usepackage{array}
\usepackage[caption=false,font=normalsize,labelfont=sf,textfont=sf]{subfig}
\usepackage{textcomp}
\usepackage{stfloats}
\usepackage{url}
\usepackage{enumitem}
\usepackage{verbatim}
\usepackage{graphicx}
\usepackage{algorithm}
\usepackage{algpseudocode}

\usepackage{booktabs} 
\usepackage{stackengine}
\usepackage{float}
\usepackage{xcolor}
\usepackage{wasysym}
\usepackage{colortbl}
\usepackage{soul}
\usepackage{multirow}
\usepackage{ragged2e}
\usepackage{hyperref}
\usepackage{comment}

\newcommand\encircle[1]{%
\tikz[baseline=(X.base)] 
 \node (X) [draw, scale=0.75, shape=circle, inner sep=0, fill=black, text=white, minimum size=0em] {\strut #1};}

\ifCLASSOPTIONcompsoc
 \usepackage[nocompress]{cite}
\else
 \usepackage{cite}
\fi

\newcommand{\xmark}{%
 \tikz[scale=0.23] {
  \draw[line width=0.7, line cap=round] (0,0) to [bend left=6] (1,1);
  \draw[line width=0.7, line cap=round] (0.2,0.95) to [bend right=3] (0.8,0.05);
 }
}

\ifCLASSINFOpdf
\else
\fi

\usepackage{todonotes}

\begin{document}

\title{SA-DS: A Dataset for Large Language Model-Driven AI Accelerator Design Generation}

\author{Deepak Vungarala$^*$,~\IEEEmembership{Student Member,~IEEE},
       Mahmoud Nazzal$^*$,~\IEEEmembership{Member,~IEEE},
      Mehrdad Morsali,~\IEEEmembership{Student Member,~IEEE}, 
      Chao Zhang,~\IEEEmembership{Member,~IEEE},
      Arnob Ghosh,~\IEEEmembership{Member,~IEEE},
      Abdallah Khreishah,~\IEEEmembership{Senior Member,~IEEE}, and
 Shaahin~Angizi,~\IEEEmembership{Senior Member,~IEEE}
 \vspace{-3em}
    \thanks{$^*$ These authors contributed equally.\\
    This work is supported in part by the National Science Foundation under Grant No. 2228028 and Semiconductor Research Corporation (SRC).} 
\IEEEcompsocitemizethanks{\IEEEcompsocthanksitem D. Vungarala, M. Nazzal, M. Morsali, A. Ghosh, A. Khreishah, and S. Angizi are with the Department of Electrical and Computer Engineering, New Jersey Institute of Technology, Newark,
NJ 07102 USA. E-mail:\{mn69, dv336, mm2772, arnob.ghosh, abadallah, shaahin.angizi\}@njit.edu.
\IEEEcompsocthanksitem C. Zhang is with the School of Computational Science and Engineering, Georgia Institute of Technology, Atlanta, GA, USA.
E-mail: chaozhang@gatech.edu.}
}

\markboth{IEEE EMBEDDED SYSTEMS LETTERS}%
{Shell \MakeLowercase{\textit{et al.}}: }

\IEEEtitleabstractindextext{%
\begin{abstract} 
In the ever-evolving landscape of Deep Neural Networks (DNN) hardware acceleration, unlocking the true potential of systolic array accelerators has long been hindered by the daunting challenges of expertise and time investment. Large Language Models (LLMs) offer a promising solution for automating code generation which is key to unlocking unprecedented efficiency and performance in various domains, including hardware descriptive code. The generative power of LLMs can enable the effective utilization of preexisting designs and dedicated hardware generators. However, the successful application of LLMs to hardware accelerator design is contingent upon the availability of specialized datasets tailored for this purpose. To bridge this gap, we introduce the \ul{S}ystolic Array-based \ul{A}ccelerator \ul{D}ata\ul{S}et (SA-DS). SA-DS comprises a diverse collection of spatial array designs following the standardized Berkeley's Gemmini accelerator generator template, enabling design reuse, adaptation, and customization. SA-DS is intended to spark LLM-centered research on DNN hardware accelerator architecture. We envision that SA-DS provides a framework that will shape the course of DNN hardware acceleration research for generations to come. SA-DS is open-sourced under the permissive MIT license at \href{https://github.com/ACADLab/SA-DS.git}{https://github.com/ACADLab/SA-DS}.
\end{abstract}

\begin{IEEEkeywords}
Systolic array design, LLM-powered hardware synthesis, accelerator architecture \vspace{-1em}
\end{IEEEkeywords} }

\maketitle
\IEEEdisplaynontitleabstractindextext
\IEEEpeerreviewmaketitle

\section{Introduction}\label{sec:introduction}
\IEEEPARstart{A}{rtificial} Intelligence (AI) has shown a remarkable potential to address complex design problems ranging from software development to drug discovery. A key advantage of AI is the significant reduction of manual effort and expertise requirements. This promising capability of AI suggests its potential in hardware design, particularly for developing specialized AI accelerators needed to keep pace with the rapid evolution of Deep Neural Networks (DNNs) \cite{chen2016eyeriss}. In DNN hardware design, the complexity and need for expert knowledge have been major limitations \cite{ren2023survey,genc2021gemmini}. 

\par
\begin{figure}[b]\vspace{-1.2em}
\begin{center}\vspace{-1em}
\includegraphics [width=1\linewidth]{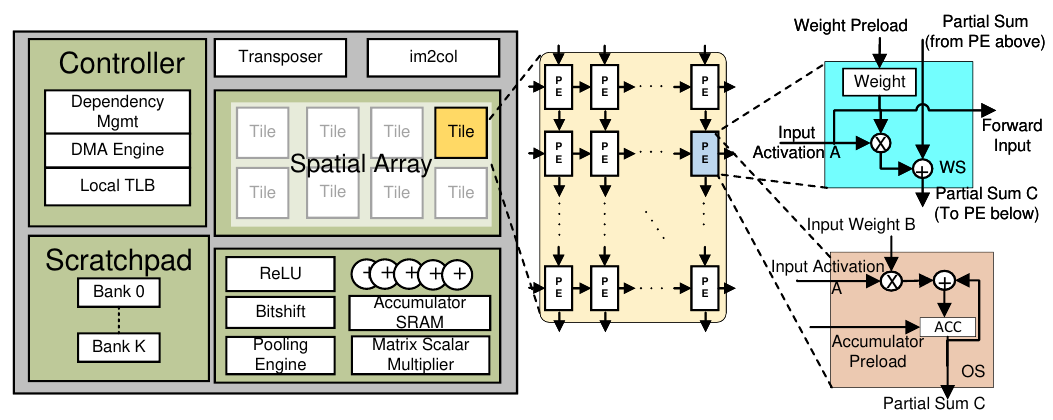}\vspace{-0.2em}
\vspace{-2em}
\caption{Gemmini architectural template for ASIC accelerator design \cite{genc2021gemmini} .}\vspace{-2.1em}
\label{Gemini}
\end{center}
\end{figure}

 A wide range of static DNN accelerator design tools have been developed such as VTA \cite{moreau2018vta}, MAGNet \cite{venkatesan2019magnet}, and DNNWeaver \cite{sharma2016high} to suit various applications. These tools provide many hardware architecture templates supporting vector/systolic spatial arrays, data flows, software ecosystems, and Operating System (OS) support. Among these tools, Gemmini \cite{genc2021gemmini} stands out as a comprehensive and well-packaged generator with open-source infrastructure tailored to designing full-stack DNN accelerators. Gemmini provides a versatile hardware framework, a multi-layered software stack, and an integrated System-on-Chip (SoC) environment based on the architectural template shown in Fig. \ref{Gemini}. At the core of Gemmini's design is a spatial array architecture, utilizing a 2-D array of tiles that contain Processing Elements (PEs). These PEs operate in parallel, efficiently handling Multiply-Accumulate (MAC) operations. 
Despite these appealing attributes, however, challenges such as the low-level nature, the complex programming interfaces, excessive memory usage, and the need for extensive development times persist \cite{xu2020automatic}. Moreover, systolic array accelerator generators like Gemmini generally face limitations in efficiently handling diverse and irregular computational patterns beyond their optimized standard operations \cite{xu2021hesa}. These limitations underscore the need for innovative solutions such as AI model-based solutions \cite{xu2020automatic,xu2021hesa}.

\par At the frontier of AI models, Large Language Models (LLMs) \cite{CHATGPT} offer an appealing solution for alleviating the challenges in hardware accelerator design.  LLMs show undoubted promise in generating HDL and High-Level Synthesis (HLS) code. Table \ref{analysis} compares notable methods in this field. GitHub Copilot \cite{friedman2021introducing} introduces automatic code generation, paving the way for tools like DAVE \cite{pearce2020dave}. Building on these efforts, VeriGen \cite{thakur2023verigen} and ChatEDA \cite{he2023chateda} refine hardware design workflows, automating the RTL to GDSII process with fine-tuned LLMs. ChipGPT \cite{chang2023chipgpt} and Autochip \cite{thakur2023autochip} integrate LLMs to generate and optimize hardware designs, with Autochip producing precise Verilog code through simulation feedback. Chip-Chat \cite{blocklove2023chip} demonstrates interactive LLMs like ChatGPT-4 in accelerating design space exploration. MEV-LLM \cite{nadimi2024multi} proposes multi-expert LLM architecture for Verilog code generation.
Assert-O \cite{miftah2024assert} is an automated framework to derive security properties from SoC documentation and optimize them.
RTLLM \cite{lu2023rtllm} and GPT4AIGChip \cite{10323953} enhance design efficiency, showcasing LLMs’ ability to manage complex design tasks and broaden access to AI accelerator design. However, most works, except GPT4AIGChip \cite{10323953}, do not address AI hardware accelerator architecture design challenges. Besides, the absence of specialized datasets for hardware accelerator design artifacts poses a significant barrier to fully harnessing the potential of LLMs \cite{he2023chateda}. This limitation confines their application to standard LLMs without fine-tuning or In-Context Learning (ICL) \cite{he2023chateda}, which are among the most promising methods for optimizing LLM capabilities \cite{dai2022can}. 

\par To bridge the gap, we introduce a Systolic Array Accelerator Dataset (SA-DS) to facilitate effective learning and generation of optimized designs by LLMs. Our contributions are as follows: (1) We create, curate, and release SA-DS, the first dataset of systolic array accelerators tailored for LLM-based DNN hardware accelerator design. Each data point in SA-DS includes a natural language description of the accelerator's micro-architecture and a Chisel description, which is an HDL representation of the design. These accelerator designs are prepared using dedicated hardware generators such as the Gemmini generator. The key advancement over existing generator-based approaches is that SA-DS offers a more effective alternative by allowing LLMs to utilize pre-existing designs instead of relying solely on hardware generators. (2) We demonstrate the potential of SA-DS in facilitating LLM-based hardware accelerator design by showcasing its effectiveness in generating viable accelerator designs using single- and multi-shot learning with multiple LLMs. Experimental results validate the suitability of SA-DS in providing high-quality and relevant accelerator design examples to contemporary LLMs, including GPT-4o \cite{openai2024gpt4o}, GPT-3.5 Turbo \cite{openai2023chatgpt}, Claude 3 Opus \cite{Claude}, and Google's Gemini \cite{Gemini}, compared to existing HDL datasets.

\begin{table*}[t]
\centering
\caption{Comparison of the state-of-the-art LLM-based HDL/HLS generators.} \vspace{-1.2em}
\scalebox{0.84}{
\begin{tabular}{|l|c|c|c|c|c|c|c|}
\hline
\rowcolor[HTML]{C0C0C0} 
\multicolumn{1}{|c|}{\cellcolor[HTML]{C0C0C0}\textbf{Function/Property}} & \textbf{Ours} & \textbf{ChatEDA \cite{he2023chateda}} & \textbf{VeriGen \cite{thakur2023verigen}} & \textbf{GPT4AIGChip \cite{10323953}} & \textbf{ChipGPT \cite{chang2023chipgpt}} & \textbf{Chip-Chat \cite{blocklove2023chip}} & \textbf{AutoChip \cite{thakur2023autochip}} \\ \hline
Function & AI Accelerator Gen.&RTL-to-GDSII&Verilog Gen.&AI Accelerator Gen.&Verilog Gen.&Verilog Gen.&Verilog Gen.
 \\ \hline
Chatbot$^*$                              & \checkmark   & \xmark          & \xmark           & \xmark              & \xmark            & \xmark            & \xmark            \\ \hline
Dataset                             & \checkmark   & NA$^\dagger$            & \checkmark(Verilog)       & \xmark              & NA            & NA            & NA             \\ \hline
Output format                                & Chisel   & GDSII           & Verilog            & HLS             & Verilog & Verilog             & Verilog            \\ \hline
Automated Verification                      & \checkmark   & \xmark$^\dagger$            & \xmark            & \xmark              & \xmark            & \xmark             & \checkmark            \\ \hline

Multi-shot examples                           & \checkmark   & \xmark            & \xmark            & \xmark              & \xmark            & \xmark             & \xmark            \\ \hline

Human in Loop                           & Low      & NA       & Medium           & Medium             & Medium           & High            & Low         \\ \hline
\end{tabular}} 
\label{analysis}

$^*$\smash[4]{A user interface featuring Prompt Optimization for the input of LLM.} 
$^\dagger$ Not applicable.
\vspace{-1.2em}
\end{table*}

\vspace{-1em}
\section{SA-DS and An Envisioned Framework} 
\par
\begin{figure}[t]
\begin{center}
\begin{tabular}{c}
\includegraphics [width=0.99\linewidth]{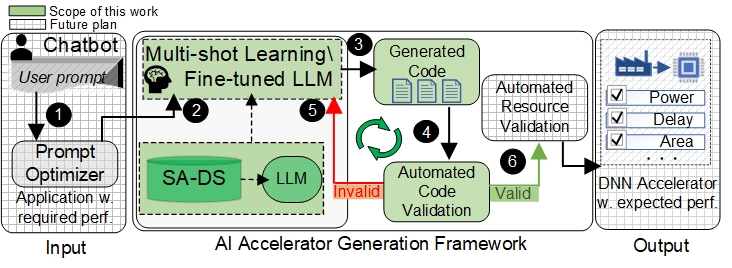}\vspace{-0.4em}
 \end{tabular} \vspace{-1.2em}\caption{An envisioned framework for utilizing SA-DS.}\vspace{-2em}
\label{Framework}
\end{center}
\end{figure}

 We propose a dataset for LLM-enhanced AI hardware accelerator design and envision a framework for its applicability. Due to space limitations, we restrict our presentation to describing how it can be utilized. This framework is outlined in Fig. \ref{Framework}. Since an LLM's response is determined by both the prompt and the model coefficients, the framework focuses on these two aspects. An immediate usage of SA-DS in the envisioned framework is to help fine-tune a generic LLM for the task of hardware accelerator design (Step \encircle{2} in Fig. \ref{Framework}). Equivalently, one may also employ ICL also known as multi-shot learning as a more computationally efficient compromise to fine-tuning \cite{dai2022can}. Besides, multi-shot prompting techniques such as ICL can be used where SA-DS will function as the source for multi-shot examples. Also, an initial prompt conveying the user's intent and key software and hardware specifications of the intended design can still be further engineered/optimized through the \textit{Prompt Optimizer} step (Step \encircle{1}). However, administering this optimization requires the development of timely and accurate evaluation techniques and metrics for the designs generated. 

\par Since SA-DS combines verbal description and systolic array design pairs, a systolic array accelerator is taken as an outcome from the LLM (step \encircle{3}). Next, a third-party quality evaluation tool can be employed to provide a quantitative evaluation of the design, verify functional correctness, and integrate the design with the full stack. (step \encircle{4}).
\par In the proposed framework, once the LLM generates systolic array accelerators (step \encircle{3}), the process moves to quality and functional evaluation (step \encircle{4}). The design, often formulated in Chisel, undergoes conversion to RTL using tools like Verilator \cite{Verilator} to assess functional correctness and full-stack integration, thus translating into a verifiable HDL. This step crucially embeds an automated RTL-GDSII validation process, where generated designs are assessed and flagged as \textit{Valid} or \textit{Invalid} based on their code sequence completeness and input-output correctness. \textit{Valid} designs advance to resource validation, focusing on optimizing for Power, Performance, and Area (PPA) metrics. Conversely, designs flagged as \textit{Invalid} trigger a feedback loop for error analysis and LLM retraining, facilitating iterative refinement (Steps \encircle{2} to \encircle{5}) to meet some preset performance criteria. Ultimately, the process culminates in Step \encircle{6}, where a script generates Tool Command Language (TCL) instructions to automate RTL evaluation for HDL codes, integrating with synthesis tools to comprehensively assess and validate the PPA metrics of the hardware design. \vspace{-0.5em}

\section{SA-DS Sample Creation}
\begin{figure}[hbt!]
\begin{center}
\begin{tabular}{c}
\includegraphics [width=0.95\linewidth,height=6.5cm]{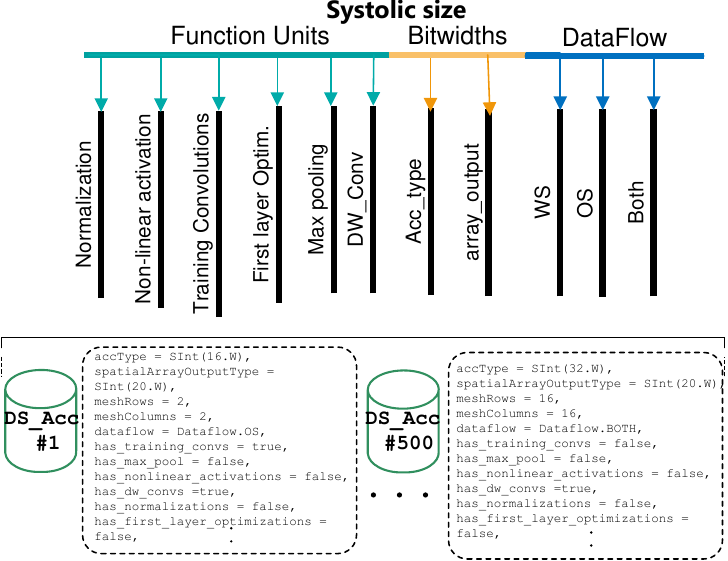}\vspace{-0.4em}
 \end{tabular} \vspace{-0.75em}
\caption{An example of one category and its design space parameters in the proposed SA-DS.} \vspace{-1.9em} 
\label{fig4}
\end{center}
\end{figure}
\vspace{-0.2em}

\par SA-DS uses the Gemmini generator to provide a variety of spatial array designs, making it easier for users to adapt and reuse these designs for different use cases. The dataset and design tools are developed with Chisel, a programming language embedded in Scala \cite{Scala}, known for its clear and efficient coding style \cite{bachrach2012Chisel}. Gemmini's configurable nature allows for significant customization, meeting various application-specific requirements, thus supporting the advancement in AI chip design \cite{chen2023diffrate}. This combination of a versatile template like Gemmini and a powerful design language like Chisel promises that SA-DS can effectively meet the diverse needs of hardware design in AI applications.

\par SA-DS is created within the Chipyard framework \cite{amid2020chipyard}, which ensures that the designs are verifiable. The process focuses on generating spatial array structures and function units from the Gemmini codebase. We refer the interested reader to the dataset's web page at \href{https://github.com/ACADLab/SA-DS.git}{https://github.com/ACADLab/SA-DS} for an algorithmic description of the steps of the generation process. The variables and their values are carefully chosen based on extensive testing with Gemmini, leading to a diverse set of potential Gemmini designs, as shown in Fig. \ref{fig4}. 

\begin{figure}[b]\vspace{-1em}
  \centering
  \includegraphics[width=0.99\linewidth]{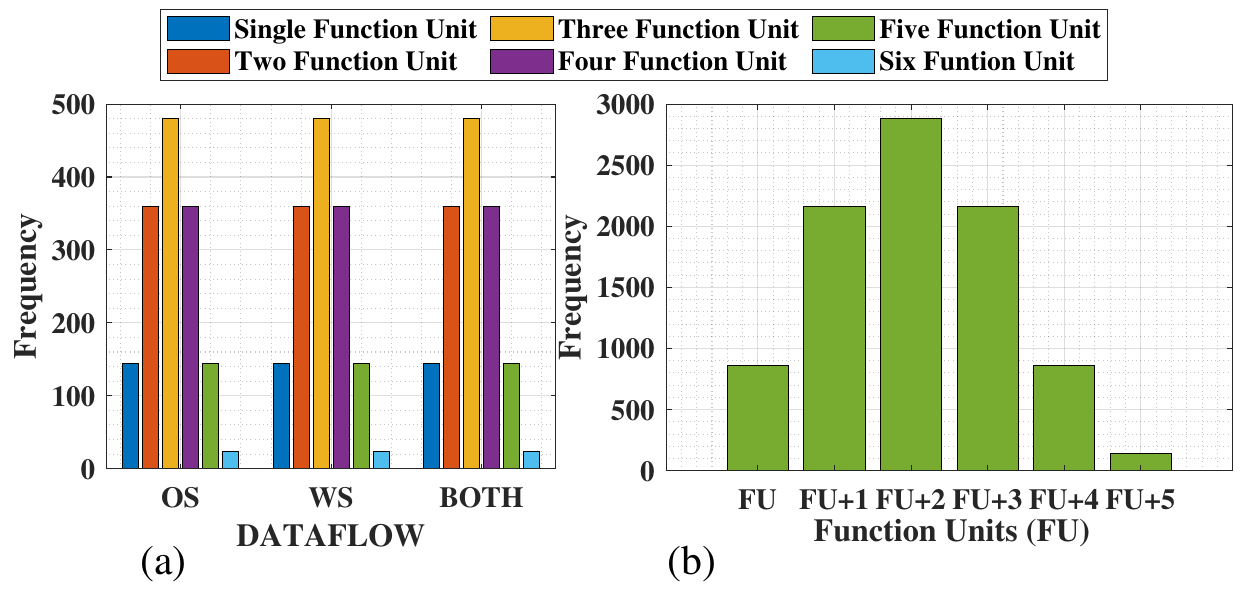} \vspace{-2.5em}
  \caption{The frequency of sample in SA-DS in terms of (a) function units in each category based on the Data flow for systolic arrays, (b) Function units available individually or in combination with the others.}
  \label{dataplot}\vspace{0em}
\end{figure}

\par SA-DS offers various configurations influenced by Function Unit (FU) availability and spatial array sizes, resulting in a structured and diverse dataset. This dataset applies to diverse hardware design needs. We carefully vary the computational and other FUs without altering the memory parameters. Due to the absence of an evaluator for obtaining design metrics, the memory parameters for any design can be either minimized or maximized based on the DNN workload. Therefore, we fixed this parameter. It is noted that when the memory parameter is fixed, Gemmini provides up to six possible design variations. Hence, we decided to exploit these six variations for each fixed memory parameter, resulting in six broad design categories. To enhance clarity, we categorized these sizes, with each category and its respective FUs depicted in Fig. \ref{fig4}. Each category includes six FU options. Given that these options are binary, there are $2^6 = 64$ possible combinations. Metrics such as input type, calculated as $64 \times 8 = 512$, represent the enriched number of data points available in each dataflow type. Considering additional dataflow variations such as Output Stationary (OS) and Weight Stationary (WS), each category encompasses $512 \times 3 = 1536$ data points as illustrated in Fig. \ref{dataplot}(a). The distribution of these parameters and their corresponding FUs is illustrated in Fig. \ref{dataplot}(b), facilitating an understanding of the dataset’s comprehensive nature and the interplay between different function units in each configuration. The overview in both dataflow and FU from Fig. \ref{dataplot} depicts a Gaussian distribution, which showcases the fairness of the dataset. \vspace{-1em}


\begin{figure} [t]
    \centering
\includegraphics[width=0.99\linewidth]{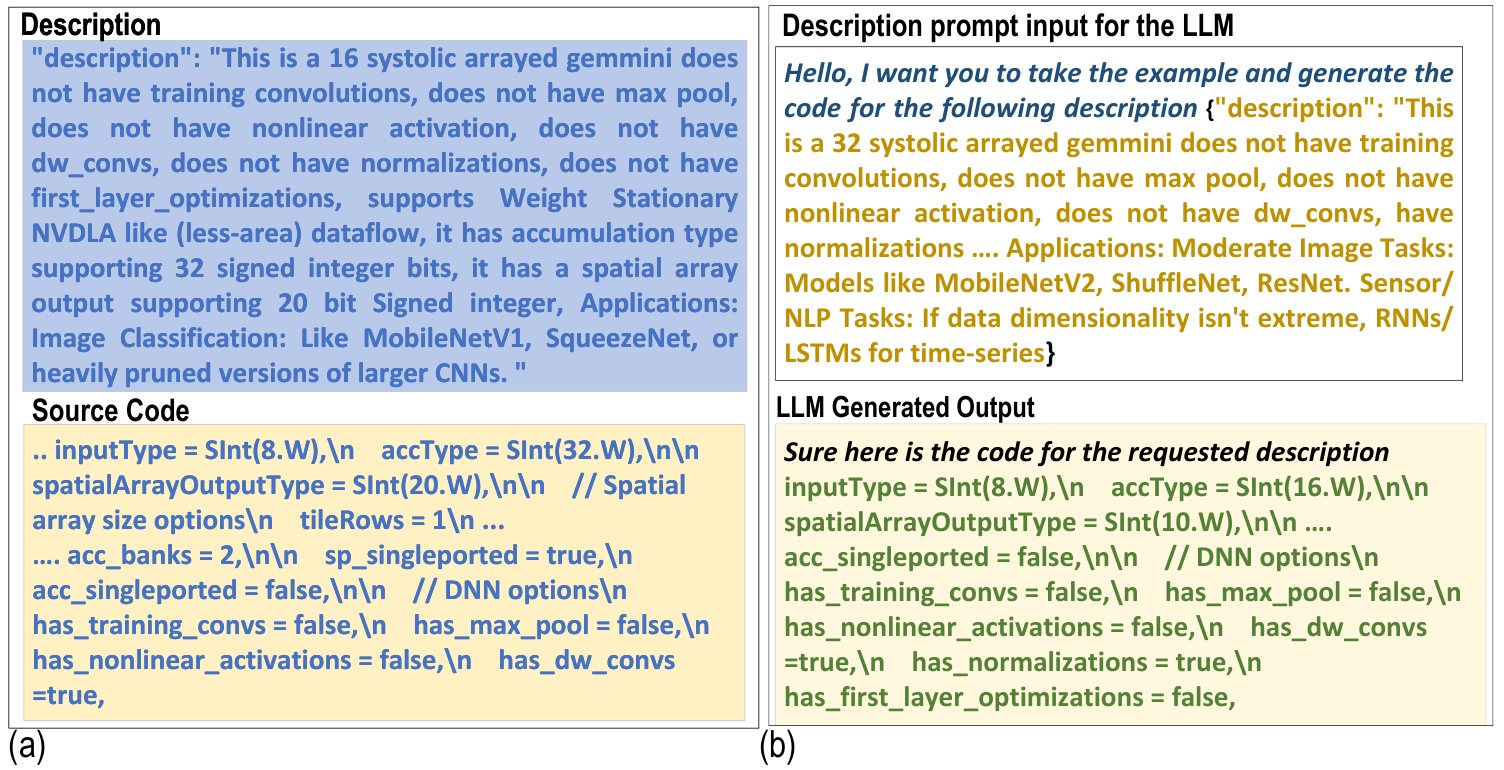} \vspace{-2em}
    \caption{(a) The data point structure, (b) An example prompt and the output generated by the LLM. }
    \label{prompt}\vspace{-1em}
\end{figure}

\section{Experimental Results}

\par In this section, we evaluate the effectiveness of SA-DS in supporting the design generation process for hardware accelerators via LLMs, referencing the framework depicted in Fig. \ref{Framework}. Due to space limitations and the complexity of LLM fine-tuning and prompt optimization, we conduct a proof-of-concept experiment comparing SA-DS with a recent HLS Dataset (HLSD) \cite{wei2023hlsdataset}\footnote{Other hardware datasets, such as \cite{lu2023rtllm}, focus on small designs like adders and multipliers, making direct comparisons unfair and difficult.}. We use each dataset to supply one-shot and multi-shot examples for LLM prompts to enhance hardware design generation from verbal descriptions. A design's verbal description is included in an instructional prompt given to the LLM for generation, as shown in Fig. \ref{prompt}. 

To objectively assess the impact of each dataset, we analyze the code quality derived from representative prompt-code pairs selected from SA-DS across four prominent LLMs: GPT-4o \cite{openai2024gpt4o}, GPT-3.5 \cite{ChatGPT-3.5}, Claude 3 Opus \cite{Claude}, and Gemini Advanced \cite{Gemini}. Reflecting the diversity of hardware specifications covered by SA-DS, for single-shot our methodology includes randomly selecting test sets from six categories within SA-DS, ensuring each category is represented by 30 samples. For multi-shot, three data points are randomly picked and queried to generate the needed by choosing the best suitable fit from the provided examples. Evaluation is conducted through a manual code review by an HLS and Chisel expert using a bi-state verification scheme. The results are summarized in Table \ref{Compar}.
A \textit{Pass} indicates the generation of complete and functional code that complies with the verbal description, or when the LLM generates the most crucial portions of the code, even if some redundant lines or extended functionalities are present. A \textit{Fail} refers to incomplete code, incorrect file headers, or fatal errors rendering the code nonfunctional. For accuracy, Verilator is used as an automated design verification tool exclusively for codes marked as \textit{Pass} for SA-DS. We focus on multi-shot learning rather than fine-tuning, as recent literature suggests that multi-shot learning is an efficient alternative to fine-tuning\footnote{Dai et al. \cite{dai2022can} report that ICL can be understood as implicit fine-tuning, demonstrating that ICL produces meta-gradients through forward computation similarly to how explicit fine-tuning updates model parameters through back-propagation.}.

\begin{table}[t]
\caption{Analysis of SA-DS vs. HLSD with single-/multi-shot prompts.} 
\resizebox{\linewidth}{!}{%
\begin{tabular}{|l|cccc|cccc|}
\hline
\multicolumn{1}{|c|}{\textbf{Input Prompt}} & \multicolumn{4}{c|}{\textbf{Single-shot}}                                & \multicolumn{4}{c|}{\textbf{Multi-shot}}                                 \\ \hline
\multicolumn{1}{|c|}{\textbf{Datasets}}     & \multicolumn{2}{c|}{\textbf{SA-DS}} & \multicolumn{2}{c|}{\textbf{HLSD}} & \multicolumn{2}{c|}{\textbf{SA-DS}} & \multicolumn{2}{c|}{\textbf{HLSD}} \\ \hline
\multicolumn{1}{|c|}{\textbf{LLMs}}         & Pass   & \multicolumn{1}{c|}{Fail}  & Pass             & Fail            & Pass   & \multicolumn{1}{c|}{Fail}  & Pass             & Fail            \\ \hline
GPT-4o                                      & 165    & \multicolumn{1}{c|}{15}    & 92               & 88              & 180    & \multicolumn{1}{c|}{0}     & 69               & 111             \\ \hline
Gemini Advanced                             & 144    & \multicolumn{1}{c|}{36}    & 57               & 123             & 180    & \multicolumn{1}{c|}{0}     & 114              & 66              \\ \hline
GPT-3.5                                     & 155    & \multicolumn{1}{c|}{25}    & 68               & 112             & 54     & \multicolumn{1}{c|}{126}   & 42               & 138             \\ \hline
Claude-3.5-sonnet                           & 150    & \multicolumn{1}{c|}{30}    & 71               & 109             & 180    & \multicolumn{1}{c|}{0}     & 57               & 123             \\ \hline
\end{tabular}%
} \vspace{-2em}
\label{Compar}
\end{table}

\par The results in Table \ref{Compar} highlight the superior performance of SA-DS over HLSD in generating hardware designs with LLMs. For single-shot prompts, SA-DS achieves significantly higher pass rates across all tested LLMs. For example, GPT-4o and Gemini Advanced show pass rates of 91.7\% and 80.0\% with SA-DS, compared to 51.1\% and 31.6\% with HLSD. Similarly, multi-shot prompts with SA-DS yield perfect pass rates (100\%) for GPT-4o, Gemini Advanced, and Claude, significantly outperforming HLSD. However, we observe a considerable and comparable fail on GPT-3.5 for both SA-DS and HLSD. 
Our observation indicates that SA-DS examples are better aligned with the LLMs’ capabilities, leading to more effective and reliable code generation. 
The higher pass rates indicate that designs generated using SA-DS typically require fewer revisions, thereby streamlining the accelerator design process. This efficiency suggests that SA-DS is highly effective in improving the overall workflow. Consequently, SA-DS showcases significant practical value by enhancing both the efficiency and quality of hardware design generation when utilizing LLMs. This reduction in necessary revisions not only saves time but also minimizes resource expenditure, underscoring the dataset's importance in AI hardware design. Furthermore, the improved quality of the initial designs can lead to faster deployment and potentially higher performance of the final products.
Besides, the higher pass rates show the SA-DS potential to solve reliability issues with 100\% success with the multi-shot approach. This underscores its practical value in streamlining the accelerator design process, making it a valuable tool for enhancing the efficiency and quality of hardware design generation.

\section{Conclusion}\vspace{-0.5em}
\par This study has introduced the first publicly accessible LLM prompt-Chisel code dataset, dubbed SA-DS. The prompt and code examples in SA-DS cover a wide variety of applications and design criteria. A proof-of-concept experiment has showcased the benefits of SA-DS in enabling the high-quality generation of hardware accelerator designs with mere verbal descriptions of novice users. This exemplifies the promising potential of enabling further research in the area of utilizing LLMs for automated hardware design generation. Key examples along this line include fine-tuning high-end LLMs for hardware design, optimized multi-shot learning, and prompt engineering serving the objectives of design efficiency in terms of execution time, hardware cost, and power consumption.

\vspace{-1em}
\small\bibliographystyle{IEEEtran}
\bibliography{IEEEabrv,./Ref}\vspace{-2em}

\begin{thebibliography}{10}
\providecommand{\url}[1]{#1}
\csname url@samestyle\endcsname
\providecommand{\newblock}{\relax}
\providecommand{\bibinfo}[2]{#2}
\providecommand{\BIBentrySTDinterwordspacing}{\spaceskip=0pt\relax}
\providecommand{\BIBentryALTinterwordstretchfactor}{4}
\providecommand{\BIBentryALTinterwordspacing}{\spaceskip=\fontdimen2\font plus
\BIBentryALTinterwordstretchfactor\fontdimen3\font minus
  \fontdimen4\font\relax}
\providecommand{\BIBforeignlanguage}[2]{{%
\expandafter\ifx\csname l@#1\endcsname\relax
\typeout{** WARNING: IEEEtran.bst: No hyphenation pattern has been}%
\typeout{** loaded for the language `#1'. Using the pattern for}%
\typeout{** the default language instead.}%
\else
\language=\csname l@#1\endcsname
\fi
#2}}
\providecommand{\BIBdecl}{\relax}
\BIBdecl

\bibitem{chen2016eyeriss}
Y.-H. Chen \emph{et~al.}, ``Eyeriss: An energy-efficient reconfigurable
  accelerator for deep convolutional neural networks,'' \emph{IEEE JSSC},
  vol.~52, no.~1, pp. 127--138, 2016.

\bibitem{ren2023survey}
W.-Q. Ren \emph{et~al.}, ``A survey on collaborative dnn inference for edge
  intelligence,'' \emph{Machine Intelligence Research}, vol.~20, no.~3, pp.
  370--395, 2023.

\bibitem{genc2021gemmini}
H.~Genc \emph{et~al.}, ``Gemmini: Enabling systematic deep-learning
  architecture evaluation via full-stack integration,'' in \emph{DAC}.\hskip
  1em plus 0.5em minus 0.4em\relax IEEE, 2021, pp. 769--774.

\bibitem{moreau2018vta}
T.~Moreau \emph{et~al.}, ``Vta: an open hardware-software stack for deep
  learning,'' \emph{arXiv preprint arXiv:1807.04188}, vol.~10, 2018.

\bibitem{venkatesan2019magnet}
R.~Venkatesan \emph{et~al.}, ``Magnet: A modular accelerator generator for
  neural networks,'' in \emph{ICCAD}.\hskip 1em plus 0.5em minus 0.4em\relax
  IEEE, 2019, pp. 1--8.

\bibitem{sharma2016high}
H.~Sharma \emph{et~al.}, ``From high-level deep neural models to fpgas,'' in
  \emph{MICRO}.\hskip 1em plus 0.5em minus 0.4em\relax IEEE, 2016, pp. 1--12.

\bibitem{xu2020automatic}
P.~Xu and Y.~Liang, ``Automatic code generation for rocket chip rocc
  accelerators,'' 2020.

\bibitem{xu2021hesa}
R.~Xu \emph{et~al.}, ``Hesa: Heterogeneous systolic array architecture for
  compact cnns hardware accelerators,'' in \emph{DATE}.\hskip 1em plus 0.5em
  minus 0.4em\relax IEEE, 2021, pp. 657--662.

\bibitem{CHATGPT}
\BIBentryALTinterwordspacing
(2023) Open ai chatgpt. [Online]. Available:
  \url{https://openai.com/research/gpt-4}
\BIBentrySTDinterwordspacing

\bibitem{friedman2021introducing}
N.~Friedman, ``Introducing github copilot: your ai pair programmer,''
  \emph{URL: https://github.
  blog/2021-06-29-introducing-github-copilot-ai-pair-programmer}, 2021.

\bibitem{pearce2020dave}
H.~Pearce \emph{et~al.}, ``Dave: Deriving automatically verilog from english,''
  in \emph{MLCAD}, 2020, pp. 27--32.

\bibitem{thakur2023verigen}
S.~Thakur \emph{et~al.}, ``Verigen: A large language model for verilog code
  generation,'' \emph{ACM TODAES}, 2023.

\bibitem{he2023chateda}
Z.~He \emph{et~al.}, ``Chateda: A large language model powered autonomous agent
  for eda,'' in \emph{MLCAD}.\hskip 1em plus 0.5em minus 0.4em\relax IEEE,
  2023, pp. 1--6.

\bibitem{chang2023chipgpt}
K.~Chang \emph{et~al.}, ``Chipgpt: How far are we from natural language
  hardware design,'' \emph{arXiv preprint arXiv:2305.14019}, 2023.

\bibitem{thakur2023autochip}
S.~Thakur \emph{et~al.}, ``Autochip: Automating hdl generation using llm
  feedback,'' \emph{arXiv preprint arXiv:2311.04887}, 2023.

\bibitem{blocklove2023chip}
J.~Blocklove \emph{et~al.}, ``Chip-chat: Challenges and opportunities in
  conversational hardware design,'' \emph{arXiv preprint arXiv:2305.13243},
  2023.

\bibitem{nadimi2024multi}
B.~Nadimi and H.~Zheng, ``A multi-expert large language model architecture for
  verilog code generation,'' \emph{arXiv preprint arXiv:2404.08029}, 2024.

\bibitem{miftah2024assert}
S.~S. Miftah, A.~Srivastava, H.~Kim, and K.~Basu, ``Assert-o: Context-based
  assertion optimization using llms,'' in \emph{Proceedings of the Great Lakes
  Symposium on VLSI 2024}, 2024, pp. 233--239.

\bibitem{lu2023rtllm}
Y.~Lu \emph{et~al.}, ``Rtllm: An open-source benchmark for design rtl
  generation with large language model,'' \emph{arXiv preprint
  arXiv:2308.05345}, 2023.

\bibitem{10323953}
Y.~Fu \emph{et~al.}, ``{GPT4AIGChip}: Towards next-generation ai accelerator
  design automation via large language models,'' in \emph{ICCAD}, 2023, pp.
  1--9.

\bibitem{dai2022can}
D.~Dai \emph{et~al.}, ``Why can gpt learn in-context? language models
  implicitly perform gradient descent as meta-optimizers,'' \emph{arXiv
  preprint arXiv:2212.10559}, 2022.

\bibitem{openai2024gpt4o}
\BIBentryALTinterwordspacing
(2024) Openai gpt-4. [Online]. Available:
  \url{https://openai.com/index/hello-gpt-4o/}
\BIBentrySTDinterwordspacing

\bibitem{openai2023chatgpt}
\BIBentryALTinterwordspacing
(2023) Openai gpt-3.5. [Online]. Available: \url{https://openai.com/chatgpt}
\BIBentrySTDinterwordspacing

\bibitem{Claude}
\BIBentryALTinterwordspacing
(2023) Anthropic. [Online]. Available: \url{https://www.anthropic.com}
\BIBentrySTDinterwordspacing

\bibitem{Gemini}
\BIBentryALTinterwordspacing
(2024) Gemini. [Online]. Available: \url{https://deepmind.google}
\BIBentrySTDinterwordspacing

\bibitem{Verilator}
\BIBentryALTinterwordspacing
(2024) Verilator. [Online]. Available:
  \url{https://github.com/verilator/verilator}
\BIBentrySTDinterwordspacing

\bibitem{Scala}
\BIBentryALTinterwordspacing
(2024) Scala. [Online]. Available: \url{https://www.scala-lang.org}
\BIBentrySTDinterwordspacing

\bibitem{bachrach2012Chisel}
J.~Bachrach \emph{et~al.}, ``Chisel: constructing hardware in a scala embedded
  language,'' in \emph{DAC}, 2012, pp. 1216--1225.

\bibitem{chen2023diffrate}
M.~Chen \emph{et~al.}, ``Diffrate: Differentiable compression rate for
  efficient vision transformers,'' in \emph{ICCV}, 2023, pp. 17\,164--17\,174.

\bibitem{amid2020chipyard}
A.~Amid \emph{et~al.}, ``Chipyard: Integrated design, simulation, and
  implementation framework for custom socs,'' \emph{IEEE Micro}, vol.~40,
  no.~4, pp. 10--21, 2020.

\bibitem{wei2023hlsdataset}
Z.~Wei \emph{et~al.}, ``Hlsdataset: Open-source dataset for ml-assisted fpga
  design using high level synthesis,'' in \emph{ASAP}.\hskip 1em plus 0.5em
  minus 0.4em\relax IEEE, 2023, pp. 197--204.

\bibitem{ChatGPT-3.5}
\BIBentryALTinterwordspacing
(2023) {ChatGPT-3.5}. [Online]. Available:
  \url{https://openai.com/blog/chatgpt}
\BIBentrySTDinterwordspacing

\end{thebibliography}

\end{document}